\documentstyle[preprint,aps]{revtex}

\begin{document}
\draft
\tighten

\title{\Large\bf A Covariant Light-Front Model of Heavy Mesons 
	within HQET}

\author{{\bf Hai-Yang Cheng, Chi-Yee Cheung, Chien-Wen Hwang 
	and Wei-Min Zhang} \\
Institute of Physics, Academia Sinica, Taipei, Taiwan 115, R.O.C.}

\date{Revised, Jan. 5, 1998}

\maketitle

\begin{abstract}
In this paper we construct a covariant light-front model of
heavy mesons within the framework of heavy quark effective 
theory (HQET). The covariant model consists of the light-front 
heavy meson bound states constructed in the heavy quark limit 
with heavy quark symmetry and  heavy quark effective theory 
as its basis. Within this  model, the Isgur-Wise function 
and decay constants in the infinite quark mass limit can be 
evaluated in a very simple and the most general 
way. The results are ensured to be consistent with heavy quark 
symmetry. From the heavy-quark-limit bound states,  we can 
further develop a systematic approach to calculate $1/m_Q$ 
corrections from the $1/m_Q$ expansion of QCD. This covariant 
model can serve as a quasi-first-principles description
of heavy meson dynamics, namely, a phenomenological covariant bound 
state in the heavy quark limit which is consistent with heavy 
quark symmetry, plus a reliable first-principles computation
of the $1/m_Q$ corrections in HQET in terms of the $1/m_Q$
expansion of the fundamental QCD theory.
\end{abstract}

\vspace{0.5in}


\newpage

\section{Introduction}
 
In the past decade, the most significant progress made in the 
QCD description of hadronic physics is perhaps in the avenue of 
heavy quark dynamics. The analysis of heavy hadron structures has
been tremendously simplified by heavy quark symmetry (HQS) proposed 
by Isgur and Wise \cite{IW89} and the heavy quark effective theory 
(HQET) developed from QCD in terms of $1/m_Q$ expansion \cite{Georgi90}. 
HQS predicts that all the heavy to heavy mesonic decay form factors 
in the infinite heavy quark mass limit are reduced to a single universal 
Isgur-Wise function, while HQET provides a systematic framework for 
studying of the symmetry-breaking $1/m_Q$ corrections (for a review, 
see \cite{Neubert94}). Applying HQS and  HQET to heavy baryons leads 
to similar simplifications to heavy baryon structures \cite{Iw91}. 
Also, combining with chiral perturbation theory, one can 
construct a chiral Lagrangian for the low-energy interaction of 
heavy hadrons with Goldstone bosons \cite{Yan92}. Moreover, in 
terms of heavy quark expansion, HQET offers a new framework to 
systematically study the inclusive decays of heavy mesons \cite{Georgi1}.

However, the general properties of heavy hadrons, namely their 
decay constants, transition form factors and structure functions
etc., are still uncalculable within QCD, even in the infinite quark 
mass limit with the utilization of HQS and HQET. Indeed, HQS itself
is unable to determine the explicit form of the Isgur-Wise function 
and it does not and cannot provide a direct description of 
nonperturbative QCD dynamics to other heavy hadron properties. 
The lattice QCD simulations \cite{lattice} permit a 
nonperturbative approach to solve the nonperturbative QCD problem, 
but so far a direct calculation with the $b$-quark is still not 
possible due to the difficulty of placing heavy particles on the 
lattice. Very recently an alternative attempt in this regard has 
been carried out from light-front QCD\cite{zhang96} but further 
investigations are needed to obtain reliable results. Hence, 
although HQS and HQET have simplified very much the heavy quark
dynamics, a complete first-principles QCD description of heavy 
hadrons is still lacking due to the unknown nonperturbative
QCD dynamics.

In most studies, the heavy hadron decay constants 
and transition form factors are usually evaluated using various 
phenomenological models, such as the constituent quark model 
\cite{Isgur89}, the MIT bag model\cite{MIT}, QCD sum rule 
\cite{Ball91}, and the light-front quark model 
\cite{Te76,Ja90,O'D94,Fa95,St95,Zo95,Le96,Gr96,Cheung97,Cheng97,De97}. 
The nonrelativistic constituent quark model and the MIT bag 
model have been widely used in the phenomenological discussion 
of hadronic structures. However, the application of these two 
models is trustworthy only for the processes with small momentum 
transfer. The problem may be overcome by the light-front quark 
model (LFQM) \cite{Te76} which has been considered as one of the best 
effective relativistic quark models in the description 
of the exclusive heavy hadron decays 
\cite{Ja90,O'D94,Fa95,St95,Zo95,Le96,Gr96,Cheung97,Cheng97,De97}. 
Its simple expression, relativistic structure and 
predictive power have made wide applications of the LFQM 
in exploring and predicting the intrinsic heavy hadron dynamics. 
However, theoretical justification of such a model within the 
framework of QCD has not been addressed in the literature.

In fact, the phenomenological LFQM is also largely limited in its
applications, due mainly to the quark model assumptions and 
its non-manifestation of covariance as a relativistic description. 
For example, it has been emphasized in the literature that various
hadronic form factors calculated in the LFQM should be extracted 
only from the plus component of the corresponding currents because 
of the dynamical dependence of other components. Also, the LFQM 
calculations for decay processes are restricted only with the zero 
momentum transfer.  As a result, one often picks up a specific 
Lorentz frame to perform such light-front calculations. These 
computations can lead to the correct results only if the 
covariance is precisely preserved within the LFQM formulation. 
We have recently attempted to calculate the decay form factors
over the whole range of momentum transfer, and realized that 
the so-called Z-diagram (a higher Fock state contribution 
beyond the LFQM) must be incorporated in order to maintain 
the covariance \cite{Cheung97}. However, with such an extension,
the simplicity of the LFQM is lost. Without including the Z-diagram 
contribution, the heavy hadronic  form factors calculated 
directly in the LFQM over the whole range of momentum transfer 
will explicitly break relativistic covariance \cite{Cheng97,De97}. Thus, 
the current approach to extract hadronic form factors in the 
LFQM formulation by choosing a specific Lorentz frame and 
then calculating a particular component (the plus component) of 
the associated current matrix element may not be unique and 
may cause some inconsistent and even misleading results. 

To resolve the above-mentioned problems in the LFQM and to improve
the current understanding of the QCD analysis of heavy hadrons, 
we construct in this paper a covariant light-front model.
This model consists of a heavy meson bound state in the heavy 
quark limit, which is phenomenological in nature at the present time 
but is fully consistent with HQS, plus a reliable approach from 
this bound state to systematically calculate the $1/m_Q$ 
corrections within HQET in terms of the $1/m_Q$ expansion of the 
fundamental QCD theory. We expect that this covariant model fully 
based on HQS and HQET
can serve as a partially phenomenological (the bound states
in the HQS limit) and partially fundamental (the $1/m_Q$ 
corrections in QCD) description of heavy mesons.

The paper is organized as follows: We will provide in Sec. II a 
general construction of light-front bound states within HQET.
To have a consistent light-front description of heavy hadrons, 
we begin with an explicit boost covariant expression of meson 
light-front bound states which can be further expressed 
in a covariant form in the heavy quark limit. We show from the 
light-front bound state equation that such bound states obey HQS. 
With the covariant bound states in the heavy quark limit, we can 
further develop a systematical approach to calculate the $1/m_Q$ 
corrections to the heavy meson energy splitting, decay constants and 
transition form factors within HQET. This will provide a 
reliable first-principles QCD analysis of the $1/m_Q$ contributions 
to heavy hadron dynamics.

For a consistent check, we evaluate in the Sec. III the 
Isgur-Wise function $\xi(v\cdot v')$ in a complete covariant way 
so that no specific Lorentz frame has to be pre-fixed, and it is not
necessary to only calculate the plus component of the current in 
matrix elements. We can straightforwardly show that in the covariant
model the Isgur-Wise function is renormalized to unity at the 
zero-recoil point without using the detailed form of the
 wavefunction, as predicted by HQS. The general
result of the Isgur-Wise function is found as a universal
function of heavy meson decays. We also
evaluate the heavy hadron decay constants $F_P$ 
and $F_V$ in the heavy quark limit in a covariant way. Again, a  
consistent result with HQS is obtained. 

In Sec. IV we further examine the covariant requirement for the 
light-front wave functions and calculate the Isgur-Wise 
function and decay constants. In Sec. V we evaluate for the 
first time the heavy mesons mass splitting in the covariant model. 

We can see that the covariant light-front model has 
overcome the theoretical difficulties encountered in the 
conventional light-front 
formulation for heavy hadrons, removed the ambiguities in 
previous calculations, largely simplified the procedure 
of the non-covariant light-front formulation currently used 
in the literature, and further provided a first-principles
QCD analysis of the $1/m_Q$ corrections within HQET. 

\section{A Covariant Light-Front Model in HQET}

The general form of the phenomenological light-front hadronic bound 
states that are currently employed in the literature has a similar 
structure as those in the constituent quark model and preserves 
a fully relativistic description. Hence these states are often called 
the relativistic quark model or the light-front quark model. However, 
almost all the previous investigations have not carefully treated
the Lorentz structure of these bound states and paid enough attention
to the consistency with heavy quark symmetry and  heavy quark 
effective theory. As it is well known, a great advantage 
of the light-front bound states, which offer a fully relativistic 
description, is the simply manifested boost transformation in these 
bound states. To correctly describe the heavy quark 
dynamics, one should also ensure that the light-front description 
can reproduce all the results associated with HQS. Also, to 
accurately extract the heavy hadron form factors, the covariant 
problem of the light-front description of heavy hadron bound states 
must be considered. Furthermore, incorporating HQET 
with the phenomenological bound states in the heavy quark limit to 
analyze the $1/m_Q$ correlations is strongly desired. 
These are the main issues we shall address in the development of 
our covariant model. 

\subsection{Boost invariant form}
 
To make the boost symmetry manifest, we begin with the mesonic
bound states that are written in a form of exhibiting explicitly
the boost covariance,
\begin{eqnarray}   \label{lfmbs}
  |P^+,P_{\bot},S,S_z\rangle &=& \int [d^3p_1] [d^3p_2] 2(2\pi)^3 
	P^+ \delta^3(P-p_1-p_2) \nonumber \\
		& & ~~~~~~~~~~~~~~~~\times \sum_{\lambda_1,\lambda_2}
	\Psi^{SS_z}(x,\kappa_{\bot},\lambda_1,\lambda_2)
	|p_1,\lambda_1;p_2,\lambda_2 \rangle \, ,
\end{eqnarray}
where we have used a boost invariant measure and the corresponding boost 
invariant light-front $\delta$-function:
\begin{equation} 
	[d^3p]\equiv {dp^+ d^2p_\bot \over 2(2\pi)^3 p^+}~~,  
	~~~~ 2(2\pi)^3 p^+\delta^3(p-p') \equiv 2(2\pi)^3 p^+ 
		\delta(p^+-{p'}^+) \delta^2(p_\bot-p'_\bot) .
\end{equation}  
In Eq.~(\ref{lfmbs}), $p_1, p_2$ and $\lambda_1, \lambda_2$ are the 
momenta and helicities of the valence quark and antiquark, respectively, 
carried in mesons, and $\Psi^{SS_z}$ is the amplitude of the 
corresponding valence $q\bar{q}$ sector (and is simply called 
the light-front wave function) which depends only on the
longitudinal momentum fraction $x$ of the valence antiquark and 
its relative transverse momentum $\kappa_{\bot}$ with respect to 
the center-of-mass frame of the hadron,
\begin{equation}
  x = \frac{\,p_2^+\,}{\,P^+\,} \, , 
	\quad \quad \kappa_{\bot}=p_{2\bot}-x P_{\bot} \, .
\end{equation}
It is obvious that $(x, \kappa_\bot)$ is boost invariant.
In general, $\Psi^{SS_z}$ can be expressed as
\begin{equation}  \label{lfwf}
  \Psi^{SS_z}(x,\kappa_{\bot}, \lambda_1,\lambda_2)={\cal F}\,
	R^{SS_z}(x, \kappa_{\bot},\lambda_1,\lambda_2) 
	\tilde{\Phi}^{SS_z} (x,\kappa^2_{\bot}) \, ,
\end{equation}
with ${\cal F}$ the flavor part of the wave function which is the 
same as that in the constituent quark model, and $R$ and $\tilde{\Phi}$
the spin and space parts, respectively, which depend on the quark 
and gluon dynamics. 
Note that $R$ is also momentum dependent due to the relativistic
feature of quarks. In phenomenological calculations, one usually 
ignores the dynamical dependence of the light-front spin so that $R$ can be
approximately expressed by the so-called Melosh matrix \cite{melosh}.  
The valence quark Fock space $|p_1,\lambda_1;p_2,\lambda_2 \rangle$ 
is given by 
\begin{equation}
	|p_1,\lambda_1;p_2,\lambda_2 \rangle = b^\dagger(p_1,
		\lambda_1) d^\dagger(p_2,\lambda_2) |0 \rangle \, ,
\end{equation}
with
\begin{equation}
	\big\{ b(p,\lambda), b^\dagger(p',\lambda') \big\}
	 = \big\{ d(p,\lambda), d^\dagger(p',\lambda')\big\}
	 = 2(2\pi)^3 p^+ \delta^3 (p-p') \delta_{\lambda \lambda'} \, ,
\end{equation}	
which is also boost invariant.

In principle, the above light-front wave functions can be
dynamically determined from the truncated light-front bound state 
equation \cite{zhang94} (which is similar to the Bethe-Salpeter 
equation in the usual equal-time frame):
\begin{equation} \label{lfbse}
  \biggl ( M_{H}^2- M_0^2\biggr)\Psi^{SS_z}(x,\kappa_{\bot},\lambda_i)
	= \int \frac{\,dx'd^2\kappa'_{\bot}\,}{2(2\pi)^3x'}\sum_{\lambda'_i}
	V_{eff}(x,\kappa_{\bot}, \lambda_i; x',\kappa'_{\bot},\lambda'_i)
	{\Psi}^{SS_z}(x',k'_{\bot},\lambda'_i) \, ,
\end{equation}
where $M_H$ is the hadron mass, $M^2_0\equiv(p_1+p_2)^2={\kappa_\bot^2 
+ m_1^2 \over 1-x}+{\kappa_\bot^2 + m_2^2 \over x}$ is the so-called (boost)
invariant mass obtained from the free energies of the constituents 
in hadrons \cite{zhang94}, and $V_{eff}$ denotes an effective two-body 
interaction kernel. Note that by ``truncating" the Fock space to 
only the valence quark states, the dominant contribution of 
higher Fock space to the bound states is described effectively by 
$V_{eff}$. Such an effective two-body interaction on the light-front 
can be derived directly from QCD, for example, by the use of the
light-front similarity renormalization group approach developed 
recently \cite{Wilson94}.

\subsection{Heavy quark limit}

In this paper, we will focus on pseudoscalar and vector heavy mesons 
(corresponding to $S=0$ and $1$, respectively) in the heavy quark  
limit. Let $P=M_H v$ in Eq.(\ref{lfmbs}), where $v^\mu~(v^2=1)$ is 
the velocity of the heavy meson and $M_H$ its mass, and rescale 
the bound states (\ref{lfmbs}) for heavy mesons by $|P^+,P_\bot,S,
S_z\rangle=\sqrt{M_H}|H(v,S,S_z)\rangle$. Then in the heavy quark limit, 
namely $m_Q \rightarrow \infty$, Eq.~(\ref{lfmbs}) can be recast into 
\begin{eqnarray}  \label{hqslfb}
  |H(v,S,S_z)\rangle &=& \int [d^3k][d^3p_q] 2(2\pi)^3v^+ \delta^3(
	\overline{\Lambda}v-k-p_q)\sum_{\lambda_Q,\lambda_q} 
	R^{SS_z} (X, \kappa_{\bot}, \lambda_Q, \lambda_q)\nonumber \\
	& & ~~~~~~~~~~~~~~~~~ \times \Phi^{SS_z} (X,\kappa^2_{\bot}) 
	b_v^\dagger(k, \lambda_Q) d_q^\dagger (p_q, \lambda_q) 
	|0\rangle \, , 
\end{eqnarray}
where $\overline{\Lambda}=M_H-m_Q$ is the so-called residual center 
mass of heavy mesons, and $k_\mu=p_Q^\mu - m_Qv^\mu$ the residual 
momentum of the heavy quark with heavy quark mass $m_Q$. Here we have 
also defined the new light-front relative momentum: 
\begin{equation}  
	0 \leq X=p_q^+ /v^+ < \infty~~, ~~~~ -\infty < \kappa_\bot = 
		p_{q\bot}-Xv_\bot < \infty \, . 
\end{equation}
Note that we use the variable $X$ (instead of the usual longitudinal 
momentum fraction $x=p_q^+/P^+$, where $P^\mu=M_H v^\mu$, or the residual 
longitudinal momentum fraction $y=p_q^+/K^+$, where $K^\mu = \overline{
\Lambda} v^\mu$ \cite{zhang96}) because it is the appropriate longitudinal 
variable appearing in the light-front heavy meson bound state equation in 
the infinite quark mass limit, see Eq.(\ref{Qqbse}). The variable $X$ 
was first introduced in \cite{cheung95} as $X=M_H x$, and it is also 
related to the residual momentum fraction $y$ by $X=\overline{\Lambda} y$. 
The operator $b_v^\dagger(k,\lambda_Q)$ creates a heavy quark with the 
residual momentum $k^\mu$ and helicity $\lambda_Q$, 
\begin{equation}
	\{ b_v (k,\lambda_Q), ~ b_{v'}^\dagger (k',\lambda'_Q) \}
		=2 (2\pi)^3 v^+ \delta_{vv'}\delta^3(k-k') 
		\delta_{\lambda_Q \lambda'_Q} ,
\end{equation}
and $[d^3k] \equiv {dk^+ d^2 k_\bot \over 2(2\pi)^3 v^+}$. 

The spin part in the above bound state can be simply constructed in 
terms of the Lorentz structure of the pseudoscalcar and vector mesons: 
$\overline{h}_v(x) i\gamma_5 q(x)$ and $\overline{h}_v(x) \gamma^\mu 
q(x)$ whose momentum structures are $R^{SS_z}$ proportional 
to $\overline{u} (v,\lambda_Q) i \gamma^5 v (p_q, \lambda_q)$ and 
$-\overline{u}(v,\lambda_Q) \epsilon \!\!\!/ v (p_q, \lambda_q)$, 
respectively, where the imaginary number $i$ and the minus sign are 
introduced by a suitable choice of the phase factor to the pseudoscalar
and vector meson bound states, $u(v,\lambda_Q)$ and $v(p_q,\lambda_q)$ 
are spinors for the heavy quark and light antiquark, 
\begin{equation}
	\sum_\lambda u(v,\lambda)\overline{u}(v,\lambda)
	   	= {1 + {\not \! v} }~~, ~~~ 
	\sum_\lambda v(p_q,\lambda) \overline{v}(p_q,\lambda) 
		= {{\not \! p}_q - m_q }, 
\end{equation}
and $\epsilon^\mu$ is the polarization vector of the vector meson.
Then the normalized $R^{SS_z}$ is given by
\begin{equation}  \label{spin}
  R^{SS_z}(X, \kappa_{\bot}, \lambda_Q, \lambda_q) 
	= \cases{ {1\over 2} \sqrt{ 1
		\over v\cdot p_q + m_q} ~ \overline u(v, \lambda_Q) 
		i\gamma^5 v(p_q, \lambda_q) ~~~~ &{\rm for}~~ S=0, \cr 
		- {1\over 2} \sqrt{ 1 \over v\cdot p_q 
		+ m_q}~ \overline u(v, \lambda_Q)
		\! \! \not \! \epsilon v(p_q, \lambda_q) 
		~~~~ & {\rm for}~~ S=1. \cr}
\end{equation}
This result is indeed the heavy quark limit of the Melosh transformation 
given in Eq.(\ref{lfwf}).

The space part $\Phi^{SS_Z}(X,\kappa_\bot^2)$ (called the light-front 
wave function) in Eq.(\ref{hqslfb}) is the heavy quark limit of $\tilde{\Phi}
^{SS_z}(x,\kappa_\bot)$ in (\ref{lfwf}) rescaled by a factor $\sqrt{M_H}$. 
The normalization of  the heavy meson bound states in the heavy quark 
limit is then given by 
\begin{equation}  \label{nmc2}
	\langle H(v',S',S'_z) |H(v,S,S_z)\rangle = 2(2\pi)^3 v^+ 
		\delta^3(\overline{\Lambda}v'-\overline{\Lambda}v) 
		\delta_{SS'} \delta_{S_zS'_z}, 
\end{equation}
which leads to the following wave function normalization condition:
\begin{equation} \label{nwf}
    \int {dX d^2\kappa_\bot \over 2(2\pi)^3 X} 
		|\Phi^{SS_z} (X, \kappa^2_\bot)|^2 = 1.
\end{equation}

\subsection{Consistency with HQS}

In principle the heavy quark dynamics is completely described by
HQET, which is given by the $1/m_Q$ expansion of the heavy quark
QCD Lagrangian: 
\begin{eqnarray} 
	{\cal L} &=& \overline{Q} (i \not \! \! D - m_Q) Q \nonumber \\
		&=& {\cal L}_0 + \sum_{n=1}^\infty \Bigg({1 \over 2m_Q} 
		\Bigg)^n {\cal L}_n  = {\cal L}_0 + {\cal L}_m\, .
		\label{hqcdl}
\end{eqnarray}
with
\begin{equation}
	{\cal L}_0 = \overline{h}_v i v \cdot D h_v 	\, , 
\end{equation}
and
\begin{equation}
	{\cal L}_n =\overline{h}_v (i\not \! D_\bot)(-iv \cdot D)^{n-1}
		(i \not \! D_\bot) h_v \, ,
\end{equation}
where $Q(x) \equiv e^{-im_Q v \cdot x}[h_v(x) + H_v(x)]$ and 
$\not \! v h_v (x)= h_v(x), \not \! v H_v(x) = - H_v(x)$, and 
$D^\mu_\bot \equiv D^\mu - v^\mu v \cdot D$.

The heavy meson bound state, Eq.~(\ref{hqslfb}), in the heavy quark 
limit is then determined by the leading Lagrangian ${\cal L}_0$ in
the above $1/m_Q$ expansion of QCD. Quantizing ${\cal L}_0$ on
the light-front by letting $h_v = h_{v+} + h_{v-}$ ($h_{v\pm} \equiv
{1\over 2} \gamma^0 \gamma^\pm h_v$), we can derive the following 
light-front bound state equation for the heavy meson 
bound state (\ref{hqslfb}): 
\begin{eqnarray}
     \Bigg\{2\overline{\Lambda} - \Big( X + {\kappa_\bot^2 + m^2_q 
	\over X} \Big) \Bigg\} && \Psi^{SS_z}(X,\kappa_{\bot})
	\nonumber \\
    && = \int {dX' d^2\kappa'_{\bot} \over 2(2\pi)^3 X'} V_{Q\overline{q}}
	(X,\kappa_{\bot}; X', \kappa'_{\bot})\Psi^{SS_z}(X',k'_{\bot}),  
	\label{Qqbse}
\end{eqnarray}
This bound state equation was also derived within the light-front HQET 
\cite{zhang96,cheung95}, where the variable $X$ is given by 
$\overline{\Lambda} y$. The effective heavy-light quark interaction 
$V_{Q\overline{q}}$ can be obtained from the leading Lagrangian 
of (\ref{hqcdl}) through the use of the light-front similarity 
renormalization group approach. Such an effective interaction is 
independent of the heavy quark's mass and spin. In other words, 
$V_{Q\overline{q}}(X,\kappa_{\bot}; X', \kappa_{\bot}')$ respects
HQS. For a detailed derivation, see \cite{zhang96}. 

It is easy to prove that the light-front wave functions discussed
above possess heavy quark symmetry. Using the normalization
property of Eq.~(\ref{spin}), Eq.~(\ref{Qqbse}) can be reduced to
\begin{eqnarray}
     \Big\{2\overline{\Lambda} -\Big(X && + {\kappa_\bot^2 + m^2_q
	\over X} \Big) \Big\} \Phi^{SS_z} (X,\kappa^2_{\bot}) 
	\nonumber \\
    && = \int {dX' d^2 \kappa'_{\bot} \over 2(2\pi)^3 X'} 
	V_{Q\overline{q}}(X,\kappa_{\bot}; X',\kappa_{\bot}') 
	{\cal M}^{SS_z}\Phi^{SS_z} (X',\kappa'^2_{\bot}) ,  \label{Qqbse1}
\end{eqnarray}
where 
\begin{equation}
	{\cal M}^{SS_z} = {1 \over 4} { 1 \over \sqrt{(v\cdot p_q 
		+ m_q)(v\cdot p'_q + m_q)}} {\rm Tr} \Big[
	(\not \! v -1) v(p,\lambda_q) \overline{v}(p',\lambda_q)\Big],
\end{equation}
which is independent of the mass and spin of the heavy quark. Thus,
the integral kernel, $V_{Q\overline{q}}(X,\kappa_{\bot}; X',
\kappa_{\bot}'){\cal M}^{SS_z}$ which determines the wave function 
$\Phi^{SS_z}$,
is also independent of the mass and spin of the heavy quark. Therefore, 
the wavefunction $\Phi^{SS_z} (X,\kappa^2_{\bot})$ must be degenerate for 
$S=0$ and $S=1$.  As a result, we can simply write
\begin{equation} \label{msiwf}
	\Phi^{SS_z}(X,\kappa^2_\bot) = \Phi(X,\kappa^2_\bot) 
\end{equation}
in the heavy quark limit. Eq.~(\ref{hqslfb}) together with Eq.~(\ref{msiwf}) 
is then the heavy meson light-front bound states in the heavy quark 
limit that obey HQS.

\subsection{The covariant form and diagrammatic rules}

The above light-front heavy mesonic bound states in the heavy 
quark limit manifest boost covariance and heavy quark 
symmetry. In fact, it can also be rewritten in a fully covariant 
form if the light-front wave function $\Phi(X,\kappa^2_\bot)$ is 
a function of $v \cdot p_q$ (see the proof given in Sec. IV): 
\begin{equation} \label{cwf}
	\Phi (X, \kappa^2_\bot) \longrightarrow \Phi (v \cdot p_q) \, ,
\end{equation}
where the antiquark $q$ in bound states is on-mass-shell, 
$p_q^- = {1\over p_q^+} (p_{q\bot}^2 + m_q^2)$. Hence, 
\begin{equation}
	v \cdot p_q = {1\over 2X} \Big( \kappa_\bot^2 + m_q^2 
		+ X^2 \Big) \, .
\end{equation}
As we will see later, the widely used Gaussian-type
wave functions have such a structure in the heavy quark limit.

When the wave function possesses such a form, the heavy meson bound
state in the HQS limit, Eq.~(\ref{hqslfb}), can be expressed 
in a covariant form:
\begin{eqnarray}  \label{hqslfbc}
  |H(v,S,S_z)\rangle = 	\overline{h}_v  \Gamma_H q_v  | 0 \rangle \, , 
\end{eqnarray}
where $\Gamma_H = i\gamma_5$ for $S=0$ and $ -\! \not \! \epsilon$ for 
$S=1$, and
\begin{eqnarray}
h_v (x) &=& \int {d^4 k \over (2\pi)^4} (2\pi) \delta(2 v \cdot k)
	\sum_{\lambda_Q} \Big\{ u(v, \lambda_Q) b_v(k,\lambda_Q) 
	e^{-ix \cdot k} + v(v, \lambda_Q) d_v^\dagger (k, \lambda_Q) 
	e^{ix \cdot k} \Big\} \, , \label{bv} \\
q_v (x) &=& \int {d^4 p_q \over (2\pi)^4} (2\pi) \delta (p_q^2 -m_q^2)
	\Phi (v \cdot p_q) {1\over 2} \sqrt{1 \over v \cdot p_q + m_q} 
		\nonumber \\
& & ~~~~~~~~~~~~~ \sum_{\lambda_q} \Big\{ u(p_q, \lambda_q) 
	b_q(p_q,\lambda_q) e^{-ix \cdot p_q} + v(p_q, \lambda_q) 
	d_q^\dagger (p_q, \lambda_q) e^{ix \cdot p_q} \Big\} \, ,
	\label{qv}
\end{eqnarray}
here $k$ is the residual momentum of the heavy quark and $p_q$ the 
momentum of the light antiquark. Note that the energies of physical 
particles in the above integrals are always positive no matter if 
they are expressed in terms of the light-front coordinates or the
equal-time coordinates. This is because we have already separated 
explicitly the particle and antiparticle components in its Fourier 
transformation. On the light-front, this implies that $n \cdot p_q 
= p_q^+ \geq 0$ [ $n$ is a light-like unit vector defined as $n^\mu 
= (0, 2, 0_\bot) $] in all the subsequent calculations presented in 
this paper. This restriction does not cause any trouble for 
maintaining the covariance.

The Fourier transformation of $\overline{h}_v(x) \Gamma_H q_v(x)$,
\begin{equation}
(\overline{h}_v \Gamma_H q_v)(v)= v \cdot n \int dx^- d^2x_\bot ~ 
  e^{-i\overline{\Lambda}v\cdot x} ~ \overline{h}_v (x) \Gamma_H 
		q_v (x) \, ,
\end{equation}
leads to Eq.~(\ref{hqslfb}), where the factor $v \cdot n$ in front 
of the Fourier transformation is due to the required normalization 
of the bound state, Eq.~(\ref{nmc2}). Now, the normalization condition of 
Eq.~(\ref{nwf}) can be written in a covariant form: 
\begin{equation} \label{cn}
	\int {d^4 p_q \over (2\pi)^4} (2\pi) \delta (p_q^2 - m_q^2)
		| \Phi ( v\cdot p_q) |^2 = 1.
\end{equation}
The left hand side of the above equation can be easily obtained in
a diagrammatic way as shown in Fig.~1:
\begin{eqnarray}
{\rm Fig.~1} &=& \int {d^4 p_q \over (2\pi)^4} (2\pi) \delta( p_q^2 
	- m_q^2) \nonumber \\
& & ~~~~~~~~~ \times |\Phi (v \cdot p_q)|^2 {1 \over 4(v \cdot p_q 
	+ m_q)}{\rm Tr} \Big[ \Gamma_H (1 + \not \! v) \Gamma_H 
		( \not \! p_q - m_q) \Big] \nonumber \\
	&=& {\rm Eq.~(\ref{cn})} \, .
\end{eqnarray}

In general, the diagrammatic rule within the covariant model is 
given as follows:

(i) The heavy meson bound state in the heavy quark limit gives a vertex
(wave function) as follows:
\begin{equation}
\begin{picture}(65,30)(0,38)
\put(20,41){\line(1,0){20}}
\put(20,39){\line(1,0){20}}
\put(40,40){\circle*{6} }
\end{picture}
	: ~~~~~~~ {1\over 2} \sqrt{1 \over v \cdot p_q + m_q} 
		~\Phi(v \cdot p_q) \Gamma_H \, ,
\end{equation}
\begin{equation}
\begin{picture}(65,30)(0,38)
\put(20,41){\line(1,0){20}}
\put(20,39){\line(1,0){20}}
\put(20,40){\circle*{6} }
\end{picture}
	: ~~~~~~~ {1\over 2} \sqrt{1 \over v \cdot p_q + m_q} 
		~\Phi^*(v \cdot p_q) \Gamma_H \, ,
\end{equation}
with a momentum conservation factor $2(2\pi)^3 ~v \cdot n ~\delta^3 
(v \overline{\Lambda}-k-p_q)$.
Here we do not have the four-momentum conservation since the meson 
and  valence quarks are all on-mass-shell, but this does not 
affect on the covariant calculations we will perform later.
 
(ii) The internal line attached to the bound state gives an on-mass-shell
propagator, 
\begin{eqnarray}
\begin{picture}(65,30)(0,38)
\put(0,40.5){\line(1,0){40}}
\put(19,40){\vector(1,0){2}}
\put(0,39.5){\line(1,0){40}}	
\put(19,28){$k$}
\end{picture}
	&:& ~~~~~~ {1 + \not \! v} ~~ ({\rm for~heavy~quarks}), ~~~ \\
\begin{picture}(65,30)(0,38)
\put(0,40){\line(1,0){40}}
\put(21,40){\vector(-1,0){2}}	
\put(17.5,30){$p_q$}
\end{picture}
	&:& ~~~~~~ { \not \! p_q - m_q} ~~ (\rm for ~light~antiquarks),
\end{eqnarray}
where $v^2=1$ and $p_q^2 = m_q^2$. 

(iii) For each internal line attached to the bound state, sum over
helicity and integrate the internal momentum using 
\begin{equation}
	\int {d^4 k \over (2\pi)^4} (2 \pi) \delta (2v \cdot k)~~~~
	{\rm and }~~~ \int {d^4 p_q \over (2\pi)^4 } (2\pi) \delta 
		(p_q^2 -m_q^2) 
\end{equation}
for the heavy and light quarks, respectively, where the delta functions 
come from the on-mass-shell condition.

(iv) For all other lines and vertices that do not attach to the bound
states, the diagrammatic rules are the same as the Feynman rules in
the conventional field theory.

These are the basic rules for the subsequent evaluations in the 
covariant model.
Now we can see that the the heavy meson states in the heavy quark limit 
consists of a bare heavy quark $h_v$ surrounded by a brown muck 
denoted by $\bar q_v$, as shown in  Fig.2. The light-front wave 
function $\Phi (v \cdot p_q)$ here can be regarded as the effective 
structure of brown muck resulted from
the complicated low-energy interactions of quarks and gluons. Therefore
the physical picture of our covariant light-front model is very clear.

Once the light-front bound states in the heavy quark limit is 
constructed, all the results obtained from the above bound states 
are ensured to be consistent with HQS. Furthermore, from such bound 
states, we can further develop a systematic approach within HQET 
to calculate the $1/m_Q$ corrections in the first-principles QCD.

\subsection{$1/m_Q$ corrections}

As we have discussed in Sec.~II.B, the heavy meson bound states in 
the infinite quark mass limit is dynamically determined by the 
leading Lagrangian ${\cal L}_0$ in the $1/m_Q$ expansion of QCD.
Then, the $1/m_Q$ corrections to the bound states can be easily
formulated in perturbation theory by taking ${\cal L}_m$ as
a perturbation in HQET. Consider the heavy meson state 
$|H(v,S,S_z)\rangle$ as a unperturbated state in the light-front
time $x^+= -\infty$. The physical state $|H(P,S,S_z) \rangle$ at
$x^+=0$ is then given by  
\begin{equation}
	|H(P,S,S_z) \rangle = ~T^+~ \exp \Bigg\{-{i\over 2} 
		\int_{-\infty}^0 dx^+ \int dx^- d^2 x_\bot ({\cal L}_m + 
		{\cal L}_{lI} )\Bigg\} | H(v,S,S_z) \rangle \, ,
\end{equation}
where $T^+$ is a light-front time-ordering operator, and ${\cal L}_{lI}$ 
the interaction part of the QCD Lagrangian for light quarks. Note that
the above light-front time-ordering perturbation expansion can be
equivalently expressed in terms of the usual equal time-ordering
perturbation expansion since the unperturbated state $|H(v,S,S_z)
\rangle$ has a covariant form.

Formally, the $1/m_Q$ corrections to physical observables are given 
as follows. Let the physical observable be $O = O_0 + O_m$, where $O_0$
is the leading operator in the $1/m_Q$ expansion, and $O_m
= \sum_{n=1}^\infty \Big({1\over m_Q}\Big)^n O_n $
is the $m_Q$-dependent terms in terms of $1/m_Q$ expansion. Then,
\begin{eqnarray}
   \langle H(P',S',S'_z)& & | O | H(P,S,S_z) \rangle  \nonumber \\
 & &= \langle H(v',S',S'_z) | T^+ \Bigg\{ O_0  + {(-i)^2 \over 2m_Q} 
	\int d^4 x_1 {\cal L}_{lI}(x_1) O_0\int d^4 x_2 
		{\cal L}_1(x_2) \nonumber \\
	& & ~~~~~~~~~~~~~~~~~~~~~~~~~~ + {(-i) \over m_Q} 
		\int d^4 x_1 {\cal L}_{lI}(x_1) O_1 + \cdots  \Bigg\} 
		| H (v,S,S_z) \rangle \nonumber \\
	& &= f_0 + {1 \over 2m_Q} f_1 + \cdots  \, .
\end{eqnarray}
Since the ``unperturbated" bound state has a covariant form, all the 
subsequent $1/m_Q$ corrections can be evaluated in a diagrammatic
way, which will be illustrated in Sec. V.

As a result, our covariant model for heavy mesons contains the bound 
state in the heavy quark limit plus a systematic approach to calculate 
the $1/m_Q$ corrections within HQET. We expect that this model can 
serve as a quasi-first-principles QCD description of heavy mesons. In
the subsequent sections, we shall give some basic applications of this 
covariant model, including the calculation of the Isgur-Wise function, 
the decay constants and the heavy meson masses up to the $1/m_Q$ 
corrections.
 
\section{A covariant evaluation of heavy meson structure in
heavy quark limit}

In this section, we shall present an extremely simple evaluation 
of the Isgur-Wise function and the decay constants in the HQS
limit within the covariant light-front model.
 
\subsection{Isgur-Wise functions}

In the infinite quark mass limit, the transition matrix elements of 
$B \rightarrow D$, and $B \rightarrow D^*$ decays are given by 
\begin{equation}
       \langle D (v') | \overline{h}_{v'}^c \Gamma
                h_v^b | B (v) \rangle ~~ {\rm and} ~~ 
		\langle D^* (v',\epsilon^*) |\overline{h}_{v'}^c 
		\Gamma h_v^b | B (v) \rangle ,     \label{5.3}
\end{equation}
where $\Gamma$ is a Dirac $\gamma$-matrix representing for the 
electroweak current. Contrary to the previous work, here we do not 
need to choose a particular Lorentz frame to calculate the above matrix
elements. Also once we have a covariant light-front model, we can 
directly evaluate the above matrix elements without restricting to 
the plus component of currents. Using the HQS-embedded covariant 
bound states, Eq.~(\ref{hqslfbc}), the hadronic matrix elements of 
$B \rightarrow D$ and $B \rightarrow D^*$ decays are given by 
(diagrammatically, see Fig.3) 
\begin{eqnarray}
        & & \langle D (v') | \overline{h}_{v'}^c \Gamma      
                h_v^b | B (v) \rangle ={\rm Tr}\Big\{ \gamma_5 
		\Big( {1+ \not{\! v}' \over 2} \Big)\Gamma \Big( 
		{1+\not{\! v} \over 2} \Big)\gamma_5 {\cal M} 
		\Big\}  \, , \label{ppd} \\
        & & \langle D^* (v',\epsilon^*) | \overline{h}_{v'}^c \Gamma 
		h_v^b | B (v) \rangle = i {\rm Tr}\Big\{\! \not{\! 
		\epsilon}^* \Big( {1 + \not{\! v}'
                \over 2} \Big) \Gamma \Big( {1+ \not{\! v} \over 2}
                \Big) \gamma_5  {\cal M} \Big\} \, ,   \label{5.5}
\end{eqnarray}
where ${\cal M}$ is the transition matrix element for the light 
antiquark (brown muck):
\begin{eqnarray} 
        {\cal M} = \int {d^4 p_q \over (2\pi)^4} (2\pi) \delta
		(p_q^2 - m_q^2) \Phi^*( v' \cdot p_q ) \Phi ( v 
		\cdot p_q ) {m_q - \not \! p_q \over \sqrt{(v\cdot 
		p_q +m_q)(v'\cdot p_q + m_q)}} \, . \label{bmk1}
\end{eqnarray}

The structure of ${\cal M}$ dictated by Lorentz invariance has the form 
\cite{Yan92}
\begin{equation}
	{\cal M} = A + B \not \! v + C \not \! v' + D \not \! 
		v \not \! v' \, . \label{bmk}
\end{equation}
This covariant decomposition allows us to easily determine the
coefficients $A, B, C, D$ in our covariant model with the results:
\begin{eqnarray}
	A &=& \int {d^4 p_q \over (2\pi)^4} (2\pi) \delta(p_q^2 - m_q^2) 
		{\Phi^*(v' \cdot p_q ) \Phi( v \cdot p_q) 
	 \over \sqrt{(v\cdot p_q +m_q)(v'\cdot p_q + m_q)}}~m_q \, , \\
	B &=& - \int {d^4 p_q \over (2\pi)^4} (2\pi) \delta(p_q^2- m_q^2)
		{\Phi^*( v' \cdot p_q) \Phi( v \cdot p_q) \over 
		\sqrt{(v\cdot p_q +m_q)(v'\cdot p_q + m_q)}} \nonumber \\
	& & ~~~~~~~~~~~~~~~~~~~~~~~~ \times {1\over 2}
		\Bigg\{ {(v+v')\cdot p_q \over (1 + v \cdot v')} +
		{(v-v')\cdot p_q \over (1 - v \cdot v')} \Bigg\} \, , \\
	C &=& - \int {d^4 p_q \over (2\pi)^4} (2\pi) \delta(p_q^2 - m_q^2)
		{\Phi^*(v' \cdot p_q ) \Phi(v \cdot p_q) \over 
		\sqrt{(v\cdot p_q +m_q)(v'\cdot p_q + m_q)}} \nonumber \\
	& & ~~~~~~~~~~~~~~~~~~~~~~~~ \times {1\over 2}
		\Bigg\{ {(v+v')\cdot p_q \over (1 + v \cdot v')} -
		{(v-v')\cdot p_q \over (1 - v \cdot v')} \Bigg\} \, , \\
	D &=& 0  \, .
\end{eqnarray}
Then Eqs.~(\ref{ppd}) and (\ref{5.5}) can be simplified as
\begin{eqnarray}
        & & \langle D (v') | \overline{h}_{v'}^c \Gamma      
                h_v^b | B (v) \rangle =\xi(v \cdot v') {\rm Tr}\Big\{ 
		\gamma_5 \Big( {1+ \not{\! v}' \over 2} \Big)\Gamma 
		\Big( {1+\not{\! v} \over 2} \Big)\gamma_5  
		\Big\}  \, , \label{ppd1} \\
        & & \langle D^* (v',\epsilon^*) | \overline{h}_{v'}^c \Gamma 
		h_v^b | B (v) \rangle = \xi(v \cdot v') {\rm Tr}\Big\{\! 
		\not{\! \epsilon}^* \Big( {1 + \not{\! v}' \over 2} \Big) 
		\Gamma \Big( {1+ \not{\! v} \over 2} \Big) i \gamma_5  
 		\Big\}  \, ,   \label{5.51} 
\end{eqnarray}
where $\xi(v \cdot v')$ is the so-called Isgur-Wise function given by
\begin{eqnarray}
	\xi (v \cdot v') &=& A - B - C + D \nonumber \\
	 &=& \int {d^4 p_q \over (2\pi)^4} (2\pi) \delta(p_q^2
                - m_q^2) \Phi^*(v' \cdot p_q) \Phi(v \cdot p_q) 
		{m_q + (v+v') \cdot p_q /(1+v \cdot v') \over 
		\sqrt{(v\cdot p_q +m_q)(v'\cdot p_q + m_q)}} \, .
		\label{iwf}
\end{eqnarray}
It is easy to check that at the zero recoil $v \cdot v'=1$, i.e., 
$v'=v$, we have 
\begin{equation}
	\xi (1) = \int {d^4 p_q \over (2\pi)^4} (2\pi) \delta(p_q^2
		- m_q^2) |\Phi( v \cdot p_q )|^2 = 1 \, ,
\end{equation}
as a result of the normalized wave function, which is model independent 
as required by HQS.

Since $v\cdot p_q = {1\over 2X}(\kappa_\bot^2 + m_q^2 + X^2)$ and
$v'\cdot p_q = {1\over 2X'}({\kappa'}_\bot^2 + m_q^2 + {X'}^2)$, 
the integration in Eq.~(\ref{iwf}) can be explicitly expressed by
\begin{eqnarray}
	\xi (v \cdot v') &=& {1 \over 1+v \cdot v'} \int 
		{dX d^2\kappa_\bot \over 2(2\pi)^3X} \Phi^*(X',
		\kappa'^2_\bot) \Phi(X, \kappa^2_\bot) \nonumber \\
	& & ~~~~~~~ \times {X'\kappa_\bot^2 + X{\kappa'}_\bot^2
		+ (X+X')(m_q^2 + XX') + 2XX'm_q (1+v \cdot v')
	\over \sqrt{XX'[\kappa_\bot^2 + (m_q+X)^2][\kappa^{'2}
		_\bot + (m_q +X')^2 ]}} ,
\end{eqnarray}
where $X' = zX, \kappa'_\bot = \kappa_\bot +X(v_\bot - zv'_\bot)$ with 
$z=v^+/{v'}^+$. To simplify the result, we can choose $v_\bot = 
v'_\bot =0$ without loss of generality, then $z$ can be related to $v 
\cdot v'$ by
\begin{equation}
	z \rightarrow z_{\pm} = v \cdot v' \pm \sqrt{(v \cdot v')^2
		- 1}~,  ~~~~~ z_+ = {1 \over z_-} \, ,
\end{equation}
where the + $(-)$ sign corresponds to $v^3$ greater (less) than $v'^3$.
It turns out 
that \begin{eqnarray} 
	\xi (v \cdot v') &=& \int {dX d^2\kappa_\bot \over 2(2\pi)^3X} 
	 ~ {2 \sqrt{z}\over 1+z}~ \Phi^*(zX,\kappa^2_\bot)\Phi(X,
		\kappa^2_\bot) \nonumber \\
		& & ~~~~~~~ \times {\kappa_\bot^2 + (m_q + X)(m_q
		 + zX) \over \sqrt{[\kappa_\bot^2 + (m_q+X)^2]
		[\kappa^2_\bot + (m_q +zX)^2 ]}} \nonumber \\
		&\equiv& \zeta(z) = \zeta(1/z)\, .
\end{eqnarray}
Note that $v^3$ greater (less) than $v'^3$ corresponds the daughter 
meson recoiling in the negative (positive) $z$ direction in the rest 
frame of the parent meson. In other words, after setting $v_\bot = 
v'_\bot=0$, the daughter meson recoiling 
in the positive and the negative $z$ directions are the only two
possible choices of Lorentz frames. The last equality of the
above equation shows that the Isgur-Wise function thus obtained 
is independent of the recoiling direction, namely, it is truly
Lorentz invariant. The above Isgur-Wise function is  an 
improvement to that given in \cite{cheung95} where the 
calculation is somewhat oversimplified. 

We also note that the above result for the Isgur-Wise function is  
the same as that extracted from the plus component of $P \rightarrow 
P$ transition matrix element at time-like $q^2$
in the non-covariant LFQM \cite{Cheng97}. 
However, in our previous work based on the LFQM \cite{Cheng97}, the 
Isgur-Wise function extracted from the plus component of $P 
\rightarrow V$ matrix element is different in its expression 
from that for $P \rightarrow P$ transition. This indicates that 
the usual LFQM utilized before is not fully consistent with 
HQS. In other words, in the usual LFQM description, 
extracting the hadronic form factors from the plus component 
of hadronic matrix elements may lead to ambiguities due to 
lack of Lorentz covariance. Therefore, the LFQM commonly used in 
the literature for the calculation of various form factors may not be 
trustworthy. Here we have shown that the Isgur-Wise function has 
exactly the same expression for  $P \rightarrow P$ and $P \rightarrow V$ 
transitions in a simple covariant description. Our covariant model
removes possible ambiguities in the usual LFQM calculations.

\subsection{Decay constants in the heavy quark limit}

To further examine the covariant model, we shall evaluate in this 
section the decay constants in the heavy quark limit.

The decay constants of pseudoscalar and vector mesons are defined 
by $\langle 0 | A^\mu | P \rangle = i f_P p^\mu$ and $\langle 0 | 
V^\mu | P^* \rangle = f_V M_V \epsilon^\mu$, where $A^\mu = \overline{q}
\gamma^\mu \gamma_5 Q$ and $V^\mu =\overline{q}\gamma^\mu Q$, with $q$ 
and $Q$ the light and heavy quark field operators, respectively,
\begin{eqnarray}
q(x) &=& \int {d^4 p_q \over (2\pi)^4} (2\pi) \delta (p_q^2 -m_q^2)
		\sum_{\lambda_q} \Big\{ u(p_q, \lambda_q) 
		b_q(p_q,\lambda_q) e^{-ix \cdot p_q} \nonumber \\
& & ~~~~~~~~~~~~~~~~~~~~~~~~~~~~~~~~~~~~~~ + v(p_q, \lambda_q) 
	d_q^\dagger (p_q, \lambda_q) e^{ix \cdot p_q} \Big\} \, ,  
		\label{qep} \\
	Q(x) &=&  e^{-im_Qv \cdot x} h_v + {\cal O}(1/m_Q) \, ,
\end{eqnarray}
and the Fourier transformation of $h_v$ in the momentum space is given 
by Eq.~(\ref{bv}). Also note that in Eq.(\ref{qep}), $n\cdot p_q =p_q^+ 
\geq 0$. In the infinite quark mass limit, the decay constants have the 
expressions 
\begin{equation}
	\langle 0 | \overline{q} \gamma^\mu \gamma_5 h_v | P(v) 
		\rangle = i F_P v^\mu~~,
		~~~ \langle 0 | \overline{q}\gamma^\mu h_v | 
		P^*(v,\epsilon) \rangle = F_V \epsilon^\mu \, ,
\end{equation}
so that 
\begin{equation}
	F_P = f_P \sqrt{M_P}~~,~~~ F_V = f_V \sqrt{M_V} \, .
\end{equation}
HQS demands that $F_V=F_P$.

Now, using the bound states in Sec.II, it is very simply to evaluate 
the above matrix elements (diagrammatically shown in Fig.~4):
\begin{eqnarray}
\langle 0 | \overline{q} \gamma^\mu \gamma_5 h_v |P (v) \rangle 
	&=& - i {\rm Tr}\Big\{\gamma^\mu \gamma_5 {1 + \not \! v 
		\over 2} \gamma_5 {\cal M}_1 \Big\} \, , \\
\langle 0 | \overline{q} \gamma^\mu h_v |P^*(v, \epsilon) \rangle 
		&=& {\rm Tr}\Big\{\gamma^\mu {1 + \not \! v 
		\over 2} \not \! \epsilon {\cal M}_1 \Big\} \, , 
\end{eqnarray}
where
\begin{equation}  \label{covint1}
	{\cal M}_1 = \sqrt{N_c} \int {d^4 p_q \over (2\pi)^4} (2\pi) 
		\delta(p_q^2- m_q^2) {\Phi(v \cdot p_q) \over 
		\sqrt{v \cdot p_q +m_q}} ( m_q - \not \! p_q) 
		= A_1 + B_1 \! \not \! v \, ,
\end{equation}
and 
\begin{eqnarray}
	A_1 &=& \sqrt{N_c} \int {d^4 p_q \over (2\pi)^4} (2\pi) 
		\delta(p_q^2- m_q^2) {\Phi( v \cdot p_q) \over 
		\sqrt{v \cdot p_q +m_q}}~ m_q ~~,  \\
	B_1 &=& - \sqrt{N_c} \int {d^4 p_q \over (2\pi)^4}(2\pi) 
		\delta(p_q^2- m_q^2) {\Phi(v \cdot p_q) \over 
		\sqrt{v \cdot p_q +m_q}} ~v \cdot p_q \, .
\end{eqnarray}
Here $N_c=3$ is the number of colors. Thus, it is easily found:
\begin{eqnarray}
F_P &=& 2 (A_1 - B_1) = 2\sqrt{N_c}\int {d^4 p_q \over (2\pi)^4} 
		(2\pi) \delta(p_q^2- m_q^2) \Phi(v \cdot p_q) 
		\sqrt{ v \cdot p_q + m_q} \nonumber \\  
    &=& 2\sqrt{N_c} \int {dXd^2 \kappa_\bot \over 2(2\pi)^3 X}
		~\Phi(X,\kappa^2_\bot) \sqrt{\kappa_\bot^2 + 
		(m_q+X)^2 \over 2 X} = F_V,	\label{decc}
\end{eqnarray}
as expected from HQS. Note that this result is different from the 
non-covariant LFQM expression given in \cite{O'D94} where again,
due to the non-covariance, the decay constant is incorrectly 
extracted from the plus component alone of the corresponding current 
matrix elements. 

\section{Numerical Test of the covariant light-front model}

In the previous sections, we have constructed a covariant 
light-front model and evaluated in a very general but 
tremendously simple way the Isgur-Wise function $\xi(v \cdot v')$ 
and the decay constants $F_P$ and $F_V$ in the heavy quark limit.
As we have pointed out, covariance requires that 
light-front wave functions be a function of $v \cdot p_q$. 
Currently, there exist several phenomenological light-front wave 
functions commonly utilized in the literature. In this
section, we shall show why a covariant light-front wave function
must be a function of $v \cdot p_q$, and see if currently used wave 
functions obey this covariant condition. Then we 
shall give a numerical calculation of the Isgur-Wise function and
decay constants with the covariant wave functions.

\subsection{Covariant light-front wave functions}

Careful readers may wonder what does covariance mean for a
bound state ? Eqs.~(\ref{hqslfbc}), (\ref{bv}) and
(\ref{qv}) are only recast in a covariant form, but the
internal momenta of the quarks carried inside mesons are 
all on-mass-shell. After the integration of the light-front
energy is performed, the bound states are apparently not
covariant any more.  We should emphasize that the validity of the
covariant conditions given by Eqs.~(\ref{bmk}) and (\ref{covint1})
depends on the specific form of light-front wave functions.
For an arbitrary light-front wave function $\Phi(X, \kappa^2_\bot)$,
whether ${\cal M}$ and ${\cal M}_1$ have the covariant structure 
shown in Eqs.~(\ref{bmk}) and (\ref{covint1}) respectively, 
depends on the specific form of light-front wave functions. We  
find that to obey Eqs.~(\ref{bmk}) and (\ref{covint1}), the 
following identity must be satisfied: 
\begin{equation}  \label{ccc}
	\int {dX d^2 \kappa_\bot \over 2 (2\pi)^3 X} \Phi (X, 
	\kappa^2_\bot) \, X = \int {dX d^2 \kappa_\bot \over 
		2 (2\pi)^3 X}  \Phi (X, \kappa^2_\bot) \, {m_q^2
		+ \kappa_\bot^2 \over X} \, .
\end{equation}
Eq.(\ref{ccc}) can be satisfied if $\Phi(X, \kappa^2_\bot)$ 
is a function of $X + {m_q^2 + \kappa_\bot^2 \over X} = 2 v \cdot 
p_q$. This leads to the covariant condition Eq.~(\ref{cwf}).

The condition (\ref{ccc}) can be easily obtained by considering
the integral involved in the calculation of Eqs.~(\ref{bmk})
and (\ref{covint1}):
\begin{equation} \label{ming}
	\int {d^4p_q \over (2\pi)^4} (2\pi) \delta (p_q^2-m_q^2)
		p^\mu_q \Phi(X, \kappa^2_\bot) = a v^\mu \, ,
\end{equation}
where the right hand side is the requirement of covariance, and
the constant $a$ is given by
\begin{equation}
	a = \int {d^4p_q \over (2\pi)^4} (2\pi) \delta (p_q^2-m_q^2)
		v \cdot p_q \Phi(X, \kappa^2_\bot) \, .
\end{equation}
Using the light-front relative momentum, the integral (\ref{ming})
for each component of $p^\mu_q$ gives
\begin{eqnarray}
& & \int {d^4p_q \over (2\pi)^4} (2\pi) \delta (p_q^2-m_q^2)
	p^+_q \Phi(X, \kappa^2_\bot) = v^+ \int {dX d^2\kappa_\bot
		\over 2(2\pi)^3 X} X \Phi(X, \kappa^2_\bot) \, , \\
& & \int {d^4p_q \over (2\pi)^4} (2\pi) \delta (p_q^2-m_q^2)
		p^i_{q\bot} \Phi(X, \kappa^2_\bot) = v^i_\bot \int {dX 
     d^2\kappa_\bot \over 2(2\pi)^3 X} X \Phi(X, \kappa^2_\bot) \, , \\
& & \int {d^4p_q \over (2\pi)^4} (2\pi) \delta (p_q^2-m_q^2)
		p^-_q \Phi(X, \kappa^2_\bot) = \int {dX d^2\kappa_\bot
	\over 2(2\pi)^3 X} \Bigg(X {v^2_\bot \over v^+} +
		{\kappa^2_\bot + m_q^2 \over X} {1\over v^+} \Bigg)
 		\Phi(X, \kappa^2_\bot) \, .
\end{eqnarray}
It is easy to check that the above integrals can have the covariant
form of (\ref{ming}) only if Eq.~(\ref{ccc}) is satisfied. Then, Eq.
(\ref{ming}) can be written in a nice covariant form,
\begin{equation}
	\int {d^4p_q \over (2\pi)^4} (2\pi) \delta (p_q^2-m_q^2)
		p^\mu_q \Phi(v \cdot p_q) = v^\mu \int {d^4p_q \over 
		(2\pi)^4} (2\pi) \delta (p_q^2-m_q^2)
		v \cdot p_q \Phi( v \cdot p_q) \, .
\end{equation}

There are several popular phenomenological light-front wave
functions that have been employed to describe various hadronic 
structures in the literature. Two of them, the Bauer-Stech-Wirbel 
(BSW) wave function 
$\Phi_{BSW}(x,\kappa^2_\bot)$ \cite{BSW} and the Gaussian-type 
wave function $\Phi_{G}(x,\kappa^2_\bot)$ \cite{Gauss}, have been
widely used in the study of heavy mesons. In the heavy 
quark limit, we denote these wave functions as follows 
\cite{cheung95,Cheng97}: 
\begin{eqnarray}
	& & \widetilde{\Phi}_{BSW}(X,\kappa^2_\bot) = \sqrt{32X} 
	\left({\pi \over \omega^2}\right) \exp\Bigg\{-{\kappa_\bot^2 
		+ X^2 \over 2\omega^2 }\Bigg\}, \\
	& & \widetilde{\Phi}_G(X,\kappa^2_\bot) = 4 \Big({\pi\over 
		\omega^2}\Big)^{3/4}\sqrt{{X^2 + m_q^2 + \kappa_\bot^2 
		\over 2X^2}} \nonumber \\
	& & ~~~~~~~~~~~~~~~~~~~~~~~~~ \times \exp\Bigg\{-{1\over 
		2\omega^2} \Big[\kappa_\bot^2 + (X/2 - (m_q^2 +
		\kappa_\bot^2)/2X)^2 \Big]\Bigg\}.
\end{eqnarray}
Since we have used the fully boost covariant form for the light-front 
bound states which are different from the previous work, the wave 
functions $\Phi(X,\kappa^2_\bot)$ in the present paper are related to
$\tilde\Phi(X,\kappa^2_\bot)$ employed in the 
previous works \cite{cheung95,Cheng97} as follows:
\begin{eqnarray}
& & \Phi_{BSW} (X,\kappa^2_\bot) = \sqrt{X}~ \widetilde{\Phi}_{BSW}
        (X,\kappa^2_\bot) = 4 \sqrt{2} \left({\pi \over \omega^2}
	\right) X \exp\Bigg\{-{\kappa_\bot^2 + X^2 \over 2\omega^2 }
		\Bigg\}, \\
& & \Phi_G(X,\kappa^2_\bot) = \sqrt{X}~ \widetilde{\Phi}_G(X,
	\kappa^2_\bot) = 4 \Big({\pi\over \omega^2}\Big)^{3/4}
	\sqrt{{X^2 + m_q^2 + \kappa_\bot^2 \over 2X}} \nonumber \\
& & ~~~~~~~~~~~~~~~~~~~~~ \times \exp\Bigg\{-{1\over 2\omega^2} 
	\Big[ \kappa_\bot^2 + (X/2 - (m_q^2 +\kappa_\bot^2)/2X)^2
		\Big]\Bigg\} \, .	\label{gslfw}
\end{eqnarray}

There is a variant of the Gaussian-type wave function which is simply 
a Gaussian function of the light-front (boost) invariant mass $M_0^2$.
Unfortunately, this wave function does not have an appropriate heavy 
quark limit \cite{Cheng97}. However, by comparison with the bound state 
equations (\ref{lfbse}) and (\ref{Qqbse}) in Sec. II, we see that 
the residual light-front (boost) invariant mass in HQS limit is given by 
the form of $ (X + {\kappa_\bot^2 + m_q^2 \over X})=2 v \cdot p_q $.
Thus, we propose here a similar Gaussian-type wave function in terms 
of the residual light-front invariant mass in the HQS limit as
follows:
\begin{equation}
	\Phi_M(X,\kappa^2_\bot) = {\cal N} \exp\Bigg\{-{1\over 2\omega}
		\Big[X + {\kappa_\bot^2 + m^2_q \over X}\Big]\Bigg\}
		= {\cal N} \exp \Bigg\{ - {v \cdot p_q \over \omega}
		\Bigg\} \, ,  \label{newlfw}
\end{equation}
where ${\cal N}$ is a normalization constant.

One can see that not all the phenomenological light-front wave 
functions have the covariant property. We found that the Gaussian wave 
function and the invariant-mass wave function can be reexpressed
as a pure function of $v \cdot p_q$. The wave 
function $\Phi_M$ has already been in this form, and $\Phi_G$ can 
be rewritten in terms of $v \cdot p_q$:
\begin{eqnarray}
\Phi_G(X,\kappa^2_\bot) &=& 4 \Big({\pi\over \omega^2}\Big)^{3/4}
	\sqrt{{X^2 + m_q^2 + \kappa_\bot^2 \over 2X}} \nonumber \\
& & ~~~~~~~~~~~~~ \times \exp\Bigg\{-{1\over 2\omega^2} \Big[ 
	(X/2 + (m_q^2 +\kappa_\bot^2)/2X)^2 - m_q^2 \Big]\Bigg\}
		\nonumber \\
	&=& 4 \Big({\pi\over \omega^2}\Big)^{3/4}
		\sqrt{ v \cdot p_q} \exp\Bigg\{-{1\over 2\omega^2} 
		\Big[ (v \cdot p_q)^2 - m_q^2 \Big]\Bigg\} \, .
\end{eqnarray}
Therefore these two wave functions preserve the Lorentz covariance 
of Eqs.~(\ref{bmk}) and (\ref{covint1}), as we have numerically 
examined from Eq.~(\ref{ccc}). However, very surprisingly, the 
commonly used BSW wave function cannot be recast as a pure function 
of $v \cdot p_q$, and it does not satisfy Eqs.~(\ref{bmk}) and 
(\ref{covint1}). Hence the BSW wave function breaks the Lorentz 
covariance. Indeed, we have already found in a recent work \cite{Cheng97} 
that there is some inconsistent problem by using the BSW wave function 
to calculate various transition form factors. Now we can understand 
why the BSW wave function gives such results inconsistent with HQS found
in Refs.\cite{Cheng97,De97}. It is  interesting to see that by 
demanding relativstic covariance, we can rule out certain types
of heavy meson light-front wave functions. 

\subsection{Numerical calculations and discussions}

Now we present some numerical results.
As we have seen the phenomenological wavefunctions have an unknown 
parameter $\omega$. This parameter is proportional to the averaged
relative momentum between the heavy and light quarks, $\sqrt{
\langle \kappa_\bot^2 \rangle}$, which corresponds to the so-called 
``Fermi motion". It is of order $\overline{\Lambda}$ and can be fixed 
by the decay constant $f_B$ and light quark masses. For example, with 
the input 
\begin{equation}
	m_q=0.25~~ {\rm GeV}~, ~~~f_B = 0.180 ~~ {\rm GeV}  \, ,
\end{equation}
we obtain the value of $\omega$ for $\Phi_M$ and $\Phi_G$ listed
in Table I. We then use these values to calculate the Isgur-Wise 
function and  its slope $\rho^2$ at the zero-recoil point (see
Table I and Fig.~5).

\vskip 0.5cm
{\small Table I. Parameter $\omega$ fitted to the decay constant
$f_B$ with $m_q=0.25$ GeV  in two different covariant light-front wave 
functions. Also shown are the results for 
$\sqrt{\langle\kappa_\bot^2\rangle}$ and the slope $\rho^2$ of the 
Isgur-Wise function at the zero-recoil point.} 
\begin{center}
\begin{tabular}{|c|c|c|c|c|}  
\hline\hline ~~ & $\omega$ (GeV) & $\sqrt{\langle \kappa_\bot^2 
\rangle}$ (GeV) & $f_B$ (GeV) & $\rho^2$ \\ \hline
$\Phi_M$ & 0.334 & 0.419 & 0.180 & 1.09 \\ \hline
$\Phi_G$ & 0.490 & 0.490 & 0.180 & 1.26 \\ \hline
\hline \end{tabular} 
\end{center}

\vspace{0.2in}

It is interesting to note that, because of the loss of covariance, 
the BSW wave function fails to give a finite result for $f_B$. This 
reinforces the statement that one cannot use a non-covariant light-front 
wave function in practical applications.
 
  From the above numerical results, we see that both the covariant
wave functions $\Phi_M$ and $\Phi_G$ give a reasonable behavior
of the Isgur-Wise function. Quantitatively, different phenomenological
wave functions give slightly different results. To get some insight of
which wave function may be more reliable, we have also looked at the 
light quark mass $m_q$ and $\omega$ dependence of $\rho^2$, where
$\rho^2$ has been measured experimentally. The results 
show that for both  $\Phi_M$ and $\Phi_G$ wave functions, the
slope $\rho^2$ of the Isgur-Wise function increases with a decrease
of the averaged relative momentum $|\kappa_\bot|$, but the 
response of $\rho^2$ to $|\kappa_\bot|$ is more sensitive for
$\Phi_G$ than that for $\Phi_M$. Furthermore, the current experimental
analysis gives $\rho^2 \sim 1.0 $ \cite{cleo} which seems to favor 
the wave function 
$\Phi_M$. Of course, there are many other possible wave functions and 
the real solution should be determined from Eq.~(\ref{Qqbse}) which 
deserves further investigation.

\section{$1/m_Q$ corrections to heavy meson masses}

In this section, we present an example, the mass splitting of
pseudoscalar and vector mesons, to show how to calculate $1/m_Q$
corrections in our covariant light-front model.

In the heavy quark limit, the heavy meson masses can be written as
\begin{equation}
	M_H = m_Q + \overline{\Lambda} \, .
\end{equation}
Here $\overline{\Lambda}$ is a constant (independent of $m_Q$) and
is of order $\Lambda_{QCD}$ and $\sqrt{\langle
\kappa_\bot^2 \rangle}$. In principle, $\overline{\Lambda}$ 
can be directly determined by solving the bound state equation in 
the heavy quark limit from the leading order Lagrangian, namely 
Eq.~(\ref{Qqbse}). This is beyond the scope of the present paper.

However, our covariant model permits a systematic evaluation of various 
$1/m_Q$ corrections, as we have formulated in Sec. II. In the infinite
quark mass limit, the pseudoscalar and vector mesons are degenerate. 
Of course, in reality they are not. Their mass difference comes mainly 
from the HQS-breaking leading $1/m_Q$ corrections.  From Sec. II, 
these $1/m_Q$ corrections can be expressed as
\begin{equation}
	M_H = m_Q + \overline{\Lambda} + {1\over 2 m_Q}(\lambda_1
		+ d_H \lambda_2 ) \, .
\end{equation}
 
\noindent In the covariant model, $\lambda_1$ and $d_H\lambda_2$ 
can be diagrammatically depicted as shown in Fig.~6:

As we can see, $\lambda_1$ is the same for both pseudoscalar and vector 
mesons, whereas $d_H\lambda_2$ gives rise to the mass splitting between
these two states. Hence, we shall only focus on the calculation
of $d_H \lambda_2$.
 From the diagrammatic rules of the covariant model, we can immediately
write down the expressions of $d_H \lambda_2$ as follows:
\begin{eqnarray}
	d_H \lambda_2 &=& {\rm Tr}\Bigg\{\Gamma_H {1+ \not \! v \over 2} 
		\sigma_{\mu \nu} {1 + \not \! v \over 2} \Gamma_H 
		M_2^{\mu \nu} \Bigg\} \, , 
\end{eqnarray}
and
\begin{eqnarray}
        M_2^{\mu \nu} &=& -g^2C_f \int {d^4 p_{1q} \over (2\pi)^4} 
		\int {d^4 p_{2q} \over (2\pi)^4} (2\pi) \delta(p_{1q}^2 
		-m_q^2)(2\pi) \delta(p_{2q}^2 -m_q^2) \nonumber \\
	& & ~~~~~~~~~~~~~ \times \sqrt{1\over (v \cdot p_{1q} + m_q) 
		(v \cdot p_{2q} + m_q)}\, \Phi(v \cdot p_{1q})
		\Phi^*( v\cdot p_{2q}) \nonumber \\
	& & ~~~~~~~~~~~~~~~~~~~~ \times q^\mu D^{\nu \nu'}\Big( (\not \! 
		p_{1q} - m_q) \gamma_{\nu'} (\not \! p_{2q} - m_q) \Big) 
		\, ,
\end{eqnarray}
where $C_f$ is  a color factor, $C_f= {N_c^2-1 \over 2N_c} = 4/3$ for 
$N_c=3$, and $D^{\mu \nu}$ is the gluon propagator, in Feynman 
gauge, $D^{\mu \nu} = {-i g^{\mu \nu} \over q^2 - m_g^2}$ with $q= 
p_{1q} - p_{2q}$, and $m_g$ is considered as an effective gluon mass 
in the low-energy domain of QCD.

Using the identity
\begin{equation}
	{1 + \not \! v \over 2} \sigma_{\mu \nu} {1 + \not \! v
		\over 2} v^\mu = 0  \, , 
\end{equation}
we obtain
\begin{equation}
M^{\mu \nu}_2 \rightarrow {1\over 4} \sigma^{\mu \nu} \lambda_2   \, ,
\end{equation}
where 
\begin{eqnarray}
	\lambda_2 &=& 4g^2C_f \int {d^4 p_{1q} \over (2\pi)^4} 
	\int {d^4 p_{2q} \over (2\pi)^4} (2\pi) \delta(p_{1q}^2 
		-m_q^2)(2\pi) \delta(p_{2q}^2 -m_q^2) {1 \over q^2 
		- m_g^2 } \nonumber \\
	& & ~~~~~~~~~~~~~ \times \sqrt{1\over (v \cdot p_{1q} + m_q)
		(v \cdot p_{2q} +m_q)}\, \Phi(v \cdot p_{1q}) 
		\Phi^*( v\cdot p_{2q})   \nonumber \\
& & ~~~~~~~~~~~~~~~~~ \times {1\over 3} \Big\{ (v \cdot p_{1q} +
	m_q) \Big[ (q \cdot p_{2q}) - (v \cdot q) (v \cdot p_{2q})
		\Big] 		\nonumber \\
& & ~~~~~~~~~~~~~~~~~~~~~~~ - (v \cdot p_{2q} + m_q)\Big[ ( q \cdot 
	p_{1q}) - (v\cdot q) (v \cdot p_{1q}) \Big] \Big\}	\, .
\end{eqnarray}
Since
\begin{equation}
	{\rm Tr}\Bigg\{ \Gamma_H {1+ \not \! v \over 2} 
		\sigma_{\mu \nu} {1 + \not \! v \over 2} \Gamma_H 
		\sigma^{\mu \nu} \Bigg\} = \cases{ 
		-12~~~~~& $\Gamma_H = i\gamma_5$, \cr 
		4 &     $\Gamma_H = - \not \! \epsilon$, \cr}
\end{equation}
we then have
\begin{equation} 
	d_H \lambda_2 = \lambda_2 \times \cases{
		-3~~~~~& S=0 ~{\rm for~pseudoscalar~mesons}, \cr 
		1 & S=1 ~{\rm for~vector~mesons} \, . \cr}
\end{equation}
This is the most general result valid in any Lorentz frame. To carry out
the numerical calculation, it is convenient to take the rest frame
$\vec{v}=0$. The numerical result for the wave function $\Phi_M$ with 
$m_b=4.8$ GeV, $m_g=0$ gives $\alpha_s = 0.29$ by fitting to the 
experimental $B - B^*$ mass splitting $\Delta M_{BB^*} \equiv 
M_{B^*}-M_B= 0.046$ GeV \cite{PDG}, and $\alpha_s=0.30$ (with $m_c
=1.6$ GeV) to the $D- D^*$ mass splitting $\Delta M_{DD^*} = 0.142$ GeV 
\cite{PDG}. For the wave function $\Phi_G$, 
we obtain $\alpha_s = 0.30$ and $0.31$ for $B-B^*$ and $D-D^*$ mass 
splittings, respectively.

\section{Conclusions and perspective}

In conclusion, we have constructed in this paper a new light-front
model for heavy mesons. The main feature of this model is the
preservation of covariance in the LFQM. We show that the covariant
light-front wave function must be subject to the constraint of
being a function of $v\cdot p_q$. As a result, we find that the
Gaussian-type light-front wave function satisfies this constraint
but the commonly used BSW wave function does not. This explains
the puzzle encountered in \cite{Cheng97} why we cannot get the
correct normalization of the Isgur-Wise function at zero recoil 
extracted from $P\to V$ transition using the BSW wave function. 
The covariant model removes some ambiguities often
occurred in the usual LFQM calculation due to lack of covariance, 
releases the restriction of light-front calculations that hadronic 
form factors can only be extracted from the matrix elements of the
plus component of the corresponding heavy quark currents at 
the zero momentum transfer.  Meanwhile, as we have shown in this 
paper, the covariant model  allows us to perform an extremely simple 
evaluation of various heavy meson properties. The typical examples 
given in this paper are the calculations of the Isgur-Wise function, 
the decay constants and the heavy-meson mass splitting which is a
$1/m_Q$ correction in HQET. Furthermore, we have also proposed a 
new covariant light-front wave function, Eq.~(\ref{newlfw}), from
the residual light-front (boost) invariant mass, which
gives good numerical results. Further applications to other
properties of heavy mesons and physical processes will be presented
in subsequent papers.

Another significance of the covariant model is that it provides a 
quasi-first-principles description of the  heavy meson dynamics.
Namely, although the heavy meson bound state in the heavy quark 
limit is still phenomenological in nature at present, it
is constrained by HQS and HQET. In the meantime, it offers a 
systematic approach to calculate the $1/m_Q$ corrections
based on the first-principles $1/m_Q$ expansion of QCD. This 
resembles very much the situation of the QCD analysis of deep 
inelastic scatterings for light quark systems, in which the low
energy dynamics (described by parton distribution functions)
is determined phenomenologically and perturbative corrections
are given in a fully first-principles way. Here the situation may
even be better since the phenomenological part is constrained by 
HQS and  HQET, whereas the nonperturbative QCD dynamics should
be much simpler in heavy quark limit, as has been discussed in 
Ref.\cite{zhang96}.

Besides many other applications,
there are some further development for this covariant light-front 
model. One is the extension to heavy baryons. The other is to 
solve the bound state equation, (\ref{Qqbse}), in the heavy 
quark limit in the low energy HQET.
Of course, once this equation is solved, the covariant model
constructed in this paper is no longer a model description, 
it will become a fully QCD description of heavy hadrons. We 
shall investigate these problems in the near future. 

\acknowledgements

This work is supported in part by the National Science Council of
the Republic of China under Grants NSC87-2112-M001-048,
NSC87-2112-M001-002 and NSC86-2816-M001-008L.


\begin{figure}
\caption[]{A diagrammatic form of the heavy meson state 
normalization.} 
\end{figure}
\begin{figure}
\caption[]{A schematic picture of a heavy meson state.}
\end{figure}
\begin{figure}
\caption[]{The diagram for $B \rightarrow D, D^*$ transitions in 
the covariant model.}
\end{figure}
\begin{figure}
\caption[]{The diagram for $B$ and $D$ meson decays in the covariant 
model. }
\end{figure}
\begin{figure}
\caption[]{The Isgur-Wise function as a function of $v \cdot v'$
calculated in the covariant model. The dashed line is for the usual
Gaussian-type light-front wave function Eq.~(\ref{gslfw}), and the solid 
line for the Gaussian wave function Eq.~(\ref{newlfw}) in terms of the 
residual light-front invariant mass.}
\end{figure}
\begin{figure}
\caption[]{The diagrams of $1/m_Q$ corrections to heavy meson masses.}
\end{figure}

\end{document}